# MUON COLLIDER DESIGN STATUS*

Y. Alexahin, FNAL, Batavia, IL 60510, U.S.A.

*Abstract*

Muon Collider (MC) - proposed by G.I. Budker and A.N. Skrinsky a few decades ago [1, 2] - is now considered as the most exciting option for the energy frontier machine in the post-LHC era. A national Muon Accelerator Program (MAP) is being formed in the USA with the ultimate goal of building a MC at the Fermilab site with c.o.m. energy in the range 1.5-3 TeV and luminosity of $\sim 1\text{-}5\cdot 10^{34}$ cm$^{-2}$s$^{-1}$. As the first step on the way to MC it envisages construction of a Neutrino Factory (NF) for high-precision neutrino experiments. The baseline scheme of the NF-MC complex is presented and possible options for its main components are discussed.

## INTRODUCTION

As was already clear in 60s [1, 2] muons provide an intriguing alternative to electrons and positrons in TeV energy range: due to practical absence of synchrotron radiation the collider ring can be very compact fitting on existing laboratory sites, the collision energy spread is significantly smaller due to negligible beamstrahlung and can be made as small as a few units by $10^{-4}$ by applying a monochromatization scheme. Another obvious advantage is by $(m_\mu/m_e)^2$ times larger *s*-channel cross-section which makes muon collider potentially a more effective tool in search for scalar particles, such as the Higgs boson.

However, short lifetime of muons – 2.2 μsec in the rest frame – makes a muon collider very challenging technologically. In his talk at Morges seminar in 1971 [2] A.N.Skrinsky briefly outlined four major requirements to render such a machine feasible: high-intensity proton driver, efficient muon production and collection scheme (so-called front-end), ionization cooling channel and, finally, fast acceleration of muons.

In a later paper [3] devoted to various cooling techniques (including the ionization cooling) it was proposed to use cooled muon beams also as the source of neutrino beams for high-precision neutrino experiments – a concept which became later known as the Neutrino Factory. The modern look at physics possibilities at a NF and MC is presented in [4].

Since mid-90s there has been some theoretical and experimental effort in the framework of international Neutrino Factory and Muon Collider Collaboration (NFMCC) which lead to successful completion of MERIT experiment at CERN on pion production in Hg jet target [5] and launching of the Muon Ionization Cooling Experiment (MICE) now under construction at RAL [6].

A significant technological progress which was achieved during the past decade and better understanding of the underlying accelerator physics made the muon collider idea look more realistic and resulted in formation of a national Muon Accelerator Program (MAP) [7] on the basis of the American part of NFMCC and the Fermilab Muon Collider Task Force created in 2006. The goal of MAP is to provide by 2015 a Design Feasibility Study Report (DFSR) which would lay the groundwork for a full-scale project aimed at the MC construction at the Fermilab site in 2020s.

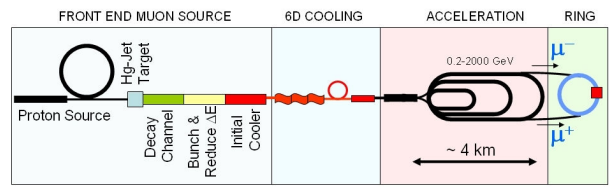

Figure 1: Schematic view of the Muon Collider complex.

## GENERAL SCHEME

A simplified scheme of a Muon Collider is shown in Fig. 1. The high-power proton beam for pion production will be provided by a chain of accelerators including those to be constructed under the Fermilab Project-X [8]. A 3 GeV 1mA CW beam from Project-X accelerators will be accumulated and re-bunched in a ring for further acceleration to 8 GeV in a pulsed linac or even up to 21 GeV if a Rapid Cycling Synchrotron option will be adopted. A possibility is also considered to accelerate the proton beam in the Main Injector up to 60 GeV to substantially reduce the required number of protons per bunch.

The accelerated proton beam should then be longitudinally compressed in another ring to be finally delivered to the pion production target with the repetition rate of the complex (10-15Hz). The pions are confined transversely by strong longitudinal magnetic field (20T at the target) lowering to 1.5-2 T in the decay channel.

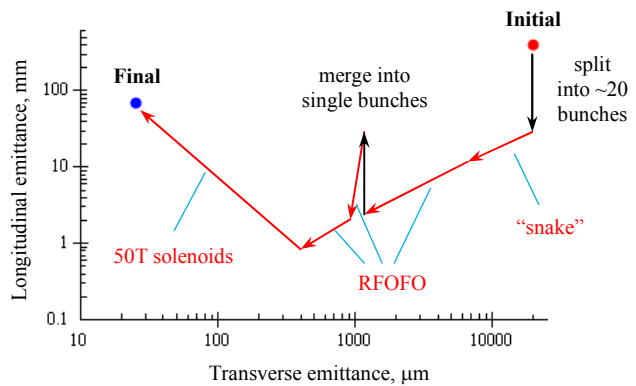

Figure 2: Evolution of muon beam emittance.

___________________________________________
* Work supported by Fermi Research Alliance, LLC under Contract DE-AC02-07CH11359 with the U.S. DOE.
#alexahin@fnal.gov

Table 1: Baseline MC parameters

| Parameter | Unit | Value |
|---|---|---|
| Beam energy | TeV | 0.75 |
| Average luminosity / IP | $10^{34}$ cm$^{-2}$s$^{-1}$ | 1.25 |
| Number of IPs, $N_{IP}$ | - | 2 |
| Circumference, $C$ | km | 2.5 |
| $\beta^*$ | cm | 1 |
| Momentum compaction, $\alpha_p$ | $10^{-5}$ | -1.5 |
| Normalized emittance, $\varepsilon_\perp$ | $\pi$·mm·mrad | 25 |
| Momentum spread | % | 0.1 |
| Bunch length, $\sigma_s$ | cm | 1 |
| Number of muons / bunch | $10^{12}$ | 2 |
| Beam-beam parameter / IP | - | 0.09 |
| RF voltage at 800 MHz | MV | 16 |
| Synchrotron tune | - | 0.0006 |
| Repetition rate | Hz | 15 |
| P-driver power | MW | 4 |

Muons produced by decaying pions are bunched in RF field with varying in time phase velocity (see next section) and then the energy of the bunches is equalised in RF rotator. The normalised r.m.s. emittance of muons captured in a bunch is ~2cm in all planes.

To cool the muons a number of steps is envisaged, the emittance evolution being plotted in Fig. 2. Both μ+ and μ- are first cooled together in a "FOFO snake" [9], then the two signs are separated and cooled individually in either "Guggenheim" RFOFO channels [10] or Helical Cooling Channels (HCC) [11] until their emittance is small enough to allow for longitudinal merging of 12-15 most populated bunches in each beam into just one bunch per beam.

After the merge 6D cooling continues until the normalised emittances reach $\varepsilon_\perp \approx 0.4$ mm, $\varepsilon_\parallel \approx 1$ mm. The final stage provides only transverse cooling while the longitudinal emittance is allowed to grow, the final values being $\varepsilon_\perp \approx 25$ μm, $\varepsilon_\parallel \approx 7$ cm.

With such longitudinal emittance, momentum spread $\sigma_p / p \approx 3\%$ and $p \approx 40$ MeV/c the bunch length will be ~ 6m so that the initial acceleration will be carried out by induction linac. After that the NF accelerating system will be used which consists of a 201 MHz linac, two RLAs and FFAG (Fig. 3). Acceleration to the final energy will be performed by a tandem of Rapid Cycling Synchrotrons (RCS).

The baseline parameters of the 1.5 TeV c.o.m. energy MC are given in Table 1.

*Neutrino Factory*

The Neutrino Factory will share the p-driver and front end with the MC. The main difference is in the packaging of protons: they will be delivered at the target in groups by 3 bunches with 50 Hz reprate. With the same average beam power the number of protons per bunch will be 10 times smaller.

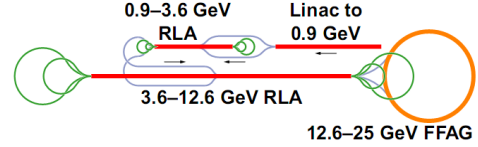

Figure 3. Layout of NF accelerators.

Another possible difference is employing of a straight FOFO channel instead of a snake: the NF does not need small longitudinal emittance so that only transverse cooling can be implemented but with acceleration of muons above 300 MeV to avoid excessive longitudinal heating.

## MAIN SYSTEMS

*Front End*

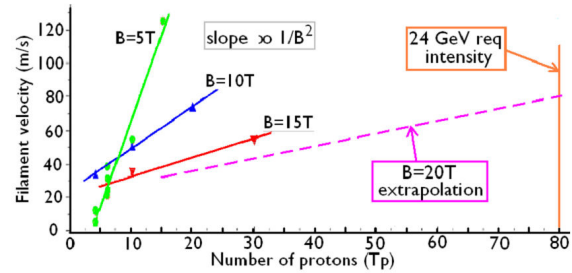

Figure 4: Observed Hg splash velocities at indicated magnetic field and their extrapolation to 20 T [12].

The use of a Hg jet as pion production target was successfully demonstrated in MERIT experiment [5]. The main issue to study was the jet explosion due to heat deposition by powerful proton beam. Fig. 4 shows the observed filament velocities vs. 24 GeV proton beam intensity and magnetic field. Projection of the measured data to 20T shows that at required proton intensities the mercury splash will be sufficiently suppressed [12].

The p-beam power which a mercury jet target can accept is estimated as 8 MW – twice the required value. There is a problem however with evacuation of energy deposited by spallation particles downstream of the target.

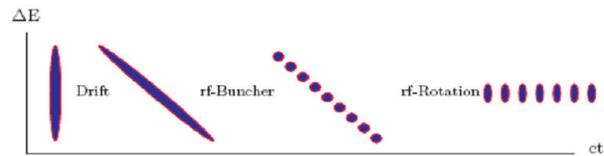

Figure 5: Muon capture in a bunch train.

In the decay channel a correlation between momentum and longitudinal position of muons is developed which is used to capture muons in wide momentum range 100-600

MeV/c [13]. The idea is illustrated by Fig. 5. RF bunching and then energy rotation is achieved with the help of RF cavities of 30 different frequencies ranging from 360 MHz at the start of the buncher to 201.25 MHz at the end of the rotator. μ+ and μ- bunches are interleaved with 180° separation in RF phase.

## 6D Ionization Cooling Channel

The major obstacle in application of ionization cooling is rapid falloff of ionization losses with particle energy leading to longitudinal heating. There are three systems under considerations with different mechanisms of the longitudinal cooling restoration.

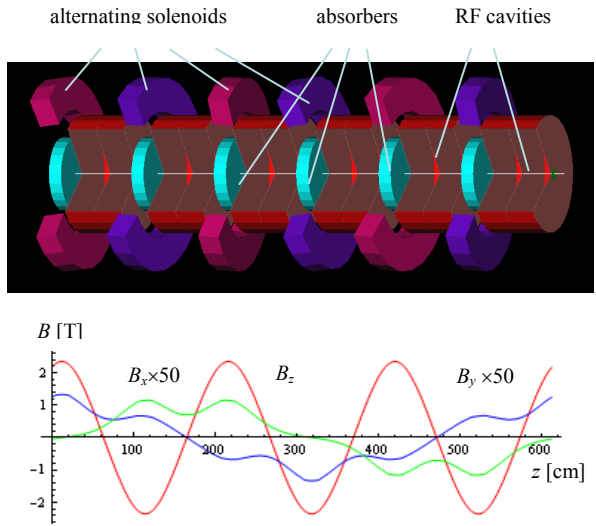

Figure 6: FOFO snake layout and magnetic field.

### FOFO snake

The first scheme – "FOFO snake" – employs dispersion in trajectory slope through a flat absorber for muons with different momenta [9]. To produce the dispersion a rotating dipole field is generated by periodically inclining the solenoids. The schematic view of one period of the channel and the magnetic field distribution along the axis are shown in Fig. 6.

Since the FOFO snake is a linear channel with flat absorbers it can cool both μ+ and μ- simultaneously. However, the amount of cooling which can be obtained in this channel is limited by relatively high beta-function value at the absorbers: 0.75 m with current design. The emittances at the snake exit – $\varepsilon_\perp \approx$ 6 mm, $\varepsilon_\parallel \approx$ 10 mm – are small enough to allow for charge separation without significant losses for subsequent cooling in RFOFO or HCC channels.

### Guggenheim RFOFO

The RFOFO (reversed FOFO) channel utilizes wedge absorbers and dispersion rather than its derivative which is created by bending the channel into a ring or a helix ("Guggenheim" RFOFO) [10]. The side view of three RFOFO cells is shown in Fig. 7. Like in the FOFO snake the solenoids have alternating polarity but owing to the unequal spacing between them the beta-function has deep minima at the absorbers – 0.4 m in the 201 MHz section – allowing to achieve smaller emittances.

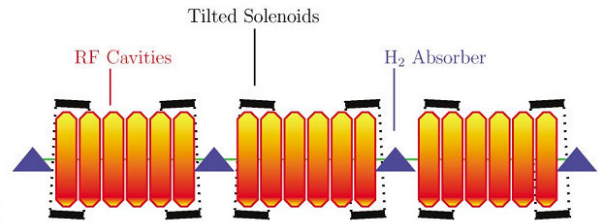

Figure 7: Schematic view of three RFOFO cells.

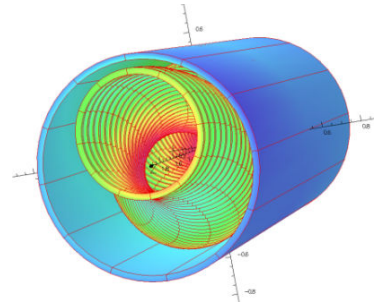

Figure 8: HCC solenoids.

### Helical Cooling Channel

The main issue with the RFOFO channel – and to lesser extent with the FOFO snake – is possible RF breakdown in strong magnetic field.

This difficulty is practically eliminated in the Helical Cooling Channel [11] which uses high-pressure $H_2$ gas filling throughout the channel as the absorber. HCC employs yet another mechanism of longitudinal cooling: large positive momentum compaction of helical orbits created by the superposition of constant longitudinal and rotating dipole fields. The right ratio of field components is obtained by using two solenoids: a helical inner solenoid and straight outer counter-solenoid (Fig. 8).

Theoretically, the existence of a continuous group of symmetry (translation + twist) makes the HCC resonance-free promising excellent dynamic properties. However, its practical implementation is quite cumbersome since the RF cavities have to be placed inside two solenoids. Another unresolved issue with HCC is RF loading with plasmas created by passing beam.

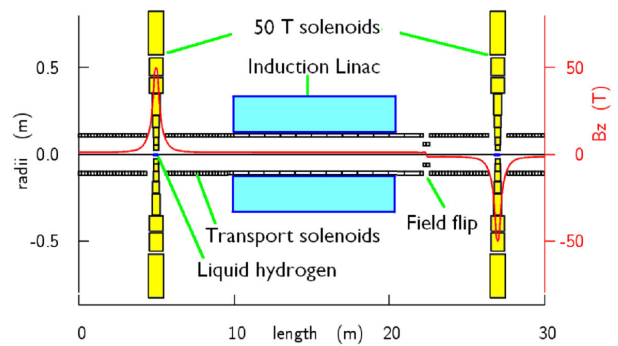

Figure 9: Concept of the 50 T solenoid chanel

## Final Cooling

To attain the required final transverse emittance, $\varepsilon_\perp \approx 25$ μm, much stronger focusing is required than can be achieved in the 6D cooling lattices. It can be obtained using high-field solenoids [12]. Progress with high temperature superconductors makes feasible magnetic fields as high as 40-50 T.

Figure 9 shows schematically a piece of such channel including two solenoids of opposite polarity. Energy lost in hydrogen absorber inside the first solenoid is replenished by induction linac which – by virtue of special waveform – also rotates the phase of the bunch so as to keep the momentum spread at minimum. The total of 13 solenoids are necessary if μ+ and μ- can be cooled in the same channel which would require the induction linac operate in bipolar regime.

Other possibilities for the final cooling are also considered including a channel with Li lenses and the so-called Parametric resonance Ionization Cooling (PIC). However, technological limitations of Li lenses and difficulties with correction of chromatic and spherical aberrations in PIC channel make them less attractive candidates than the high-field solenoids.

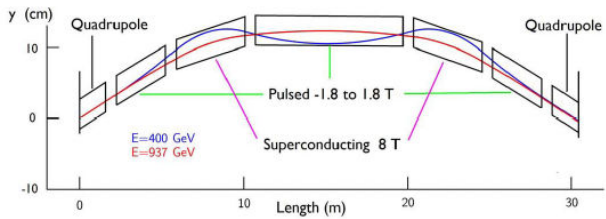

Figure 10 (color): Closed orbit through the RCS half-cell at low energy (blue) and final energy (red).

Table 2: Muon transmission for various steps.

| Step | Transmission | Cumulative |
|---|---|---|
| Best 12 bunches | 0.7 | 0.7 |
| Charge separation | 0.9 | 0.63 |
| 6D cooling before merge | 0.47 | 0.3 |
| Merge | 0.88 | 0.26 |
| 6D cooling after merge | 0.48 | 0.12 |
| Final cooling | 0.65 | 0.08 |
| Acceleration | 0.7 | 0.057 |

## Acceleration

After induction linacs at the very early stage of acceleration the MC will use the NF accelerator chain depicted in Fig. 3. For the subsequent acceleration to final energy it is planned to use Rapid Cycling Synchrotrons to maximize the number of passes through RF cavities.

To make the RCS compact and fast it is proposed to use combination of fixed-field superconducting magnets and warm AC magnets with yokes of grain-oriented silicon steel [14]. A half-cell layout is shown in Fig. 10. The University of Mississippi is building 400 Hz 1.8 T prototype magnets with tests scheduled for 2011-2012.

A RLA is also being designed (as shown in Fig. 1), but it is significantly more costly and therefore considered only as a fallback solution.

The expected efficiency of all stages of muon beam manipulations is presented in Table 2. For the front end yield of 0.2μ per 8 GeV proton the required p-driver power is 3.4 MW, so there is a good margin.

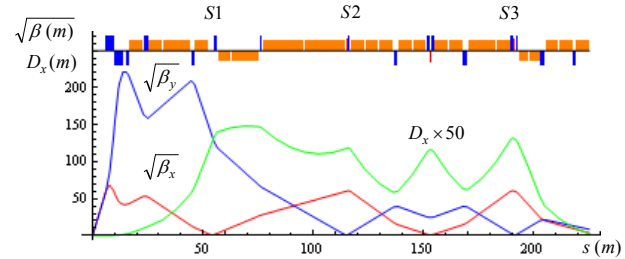

Figure 11 (color): IR layout and optics functions.

## Collider Ring

To obtain large momentum acceptance for very small $\beta^*$ a new scheme for IR chromaticity correction – called a three-sextupole scheme – was developed [15]. It includes strong dipoles in the close vicinity of IP (Fig. 11 orange boxes) to generate dispersion and at the same time to sweep decay electrons away from the detector. The design is based on the existing Nb3Sn magnet technology. MARS simulations show tolerable levels of energy deposition in magnets and detector backgrounds [16].